\documentclass{emulateapj}
\usepackage{graphicx,amssymb}
\newcommand{\kms}{\ifmmode {\rm km\,s}^{-1} \else km\,s$^{-1}$\fi}

\slugcomment{Draft Version 1.0, LOS}

\shorttitle{Methanol maser in M31}
\shortauthors{Sjouwerman et al.}

\begin{document}

\title{Discovery of the first methanol (CH$_3$OH) maser in the
  Andromeda galaxy (M31)}


\author{Lor\'ant~O.~Sjouwerman} \affil{National Radio Astronomy
  Observatory, P.O. Box 0, Lopezville Rd.\,1001, Socorro, NM 87801}

\email{lsjouwer@nrao.edu}

\author{Claire~E.~Murray} \affil{Department of Physics
  and Astronomy, Carleton College, One North College Street,
  Northfield, MN 55057}
\affil{National Radio Astronomy
  Observatory, P.O. Box 0, Lopezville Rd.\,1001, Socorro, NM 87801}

\author{Ylva~M.~Pihlstr\"om\altaffilmark{1}} \affil{Department of
  Physics and Astronomy, University of New Mexico, MSC07 4220,
  Albuquerque, NM 87131}

\altaffiltext{1}{Y.~M.~Pihlstr\"om is also an Adjunct Astronomer at
  the National Radio Astronomy Observatory}

\author{Vincent~L.~Fish} \affil{Massachusetts Institute of Technology,
  Haystack Observatory, Route 40, Westford, MA 01886}

\and 

\author{Esteban~D.~Araya} \affil{Physics Department, Western Illinois
  University, 1 University Circle, Macomb, IL 61455}

\begin{abstract}
  We present the first detection of a 6.7 GHz Class\,II methanol
  (CH$_3$OH) maser in the Andromeda galaxy (M31). The CH$_3$OH
  maser was found in a Very Large Array (VLA) survey during the fall
  of 2009. We have confirmed the methanol maser with the new Expanded
  VLA (EVLA),
  in operation since March 2010, but were unsuccessful in detecting a
  water maser at this location. A direct application for this methanol
  maser is the determination of the proper motion of M31, such as was
  obtained with water masers in M33 and IC10 previously. Unraveling
  the three-dimensional velocity of M31 would solve for the biggest
  unknown in the modeling of the dynamics and evolution of the Local
  Group of galaxies.
\end{abstract}


\section{Introduction}

The discovery of 22 GHz water masers in galaxies of the Local Group
contributed to the study of the content, dynamics and evolution of
individual star forming complexes in their host galaxies \citep[e.g.,][]{chur77,huch88,brun06a}. More
importantly, the microarcsecond ($\mu$as) accuracy provided by VLBI
observations of these masers enabled measurements of the galactic
rotation around the nucleus of their host galaxies, as well as the
proper motion of their galaxies as a whole \citep[e.g.,][]{brun06b}.  A dynamical model of the
history and mass distribution of the Local Group is within reach with
accurate knowledge of the absolute distances and three-dimensional
motions of its members.

The Milky Way and Andromeda (M31), together with M31's companion M33,
are the dominant constituents of the Local Group and are responsible
for the largest dynamic forces therein.
 Obtaining M31's geometric distance
and its transverse velocity would resolve the largest
unknown, its three-dimensional velocity, in the modeling of the Local
Group \citep[e.g.,][]{peeb01,loeb05,mare08}.

Water masers in star forming complexes are probably the brightest
masers available, and their relatively high frequency is an advantage
in proper motion studies as the accuracy scales with wavelength. On
the other hand, the higher frequency implies smaller beam sizes,
making a large area survey relatively expensive. Most often water
maser searches are targeted toward compact HII regions, where massive
star formation might be expected.

Current studies of the kinematics of the Local Group of galaxies
include measuring the transverse motions of water maser complexes in
galaxies such as M33 and IC10 with an accuracy of up to 3
$\mu$as\,yr$^{-1}$, i.e, about 10 km\,s$^{-1}$ with a time baseline of
3 years \citep{gree93,brun05,brun07}. The anticipated transverse motion of
M31 is $>$80 km\,s$^{-1}$ \citep{loeb05,mare08}, occasionally quoted
as large as $\sim$150-160 km\,s$^{-1}$
\citep{peeb94,peeb01}. Unfortunately, no water masers are found in
targeted observations toward M31 \citep[][M.\ Claussen, priv.\
comm.]{gree95,imai01} leaving a major uncertainty in the dynamical
history and mass distribution of the Local Group due to the
undetermined transverse velocity of M31.

We therefore embarked on a systematic survey for Class\,II 6.7 GHz
methanol masers in M31 to find sources for proper motion studies.
Class\,II methanol masers are radiatelively pumped and believed to be
signposts of an early phase of massive star formation.
These methanol masers represent the second brightest known Galactic
maser transition, only surpassed by 22 GHz water vapor masers.  
The astrometry at 6.7 GHz
is about a factor of three worse than at 22 GHz, in principle
requiring three times as long a time baseline in order to measure
proper motions to the same accuracy.  However, tropospheric errors are
much less problematic at 6.7 GHz than at 22 GHz, partly compensating
this disadvantage.

The detection rate for extragalactic methanol masers seems to be very
sensitive to the metallicity of the host galaxy \citep{beas96,gree08}.
Other than the Milky Way, the only other galaxy in which methanol
masers have been detected is the Large Magellanic Cloud (LMC;
\citealt[][and references therein]{gree08}). No methanol masers have
been detected in the low-metallicity Small Magellanic Cloud \citep{gree08} or
M33 \citep{gold08}. In contrast, the high metallicity of M31 makes it
an excellent target for a methanol maser search.

We will report on the full results of our systematic survey in a
forthcoming paper. Here we report on the detection of the first 6.7
GHz methanol maser in M31. We also
searched this location for a potential 22 GHz water maser.

\begin{figure*}[t]
\includegraphics[angle=0,width=\textwidth]{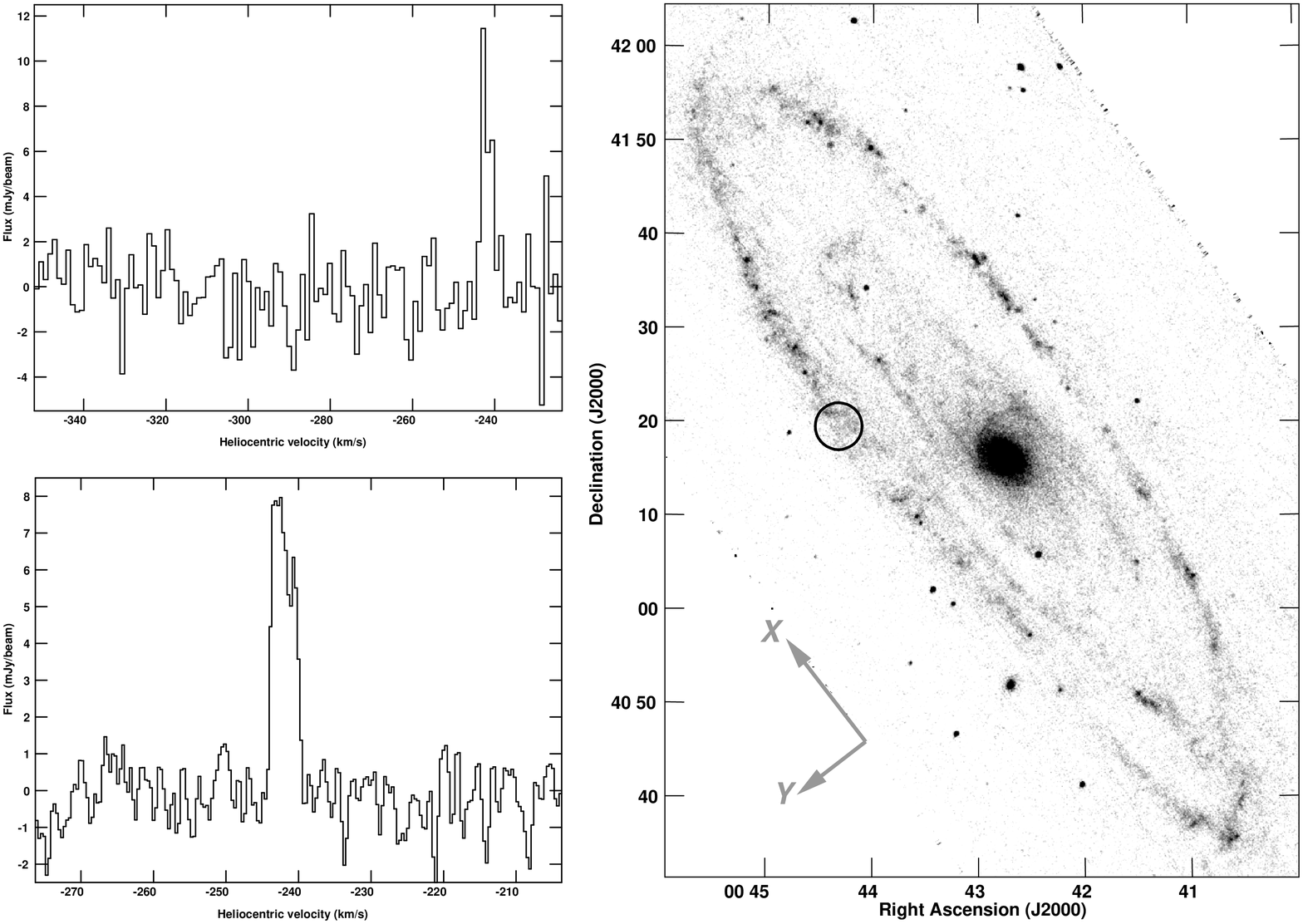}
\caption{Top left: original spectrum of the first methanol maser in
  M31, taken with the VLA correlator using EVLA receivers. Bottom
  left: Smoothed confirmation spectrum, taken using the EVLA and the
  WIDAR correlator. Right: MSX 8 $\mu$m infrared image of M31
  \citep{mosh97,krae02} locating the methanol maser in an
  inconspicuous part of the $\sim$10 kpc molecular ring (circle). The
  directions of the axes of the X,Y coordinate system (where the
  origin should be in the nucleus) of \citet{baar64} are indicated.}
\label{vladet}
\end{figure*}

\section{Observations}

The Very Large Array (VLA) was tuned to observe a rest frequency of 6668.518 MHz at
a heliocentric (optical) velocity of about $-$275 km\,s$^{-1}$. The
VLA correlator was processing a 3.125 MHz (140 km\,s$^{-1}$) Doppler
tracked dual polarization bandwidth divided in 128 frequency channels
of 24.4 kHz (1.1 km\,s$^{-1}$) each. Typical on-source times were about 2
hours per pointing, yielding typically 2.7 mJy\,beam$^{-1}$ rms noise
per frequency channel over most of the band. Over the lower $\sim$0.5
MHz, the noise was exponentially increased to about twice this value
due to aliasing effects caused by feeding EVLA receiver electronics
into the VLA correlator.  The observations took place during a
period of transition from the VLA to the Expanded VLA (EVLA), with the
result that the number of telescopes with 4--6 GHz C band receivers
varied by epoch. Also the number of new polarizers deployed on these
receivers varied per observation.  However, no fewer than 16 out of 27
possible receivers (by late 2010) were used for any of the
observations. Several candidate masers with peaks over 5$\sigma$ were
cataloged.

We obtained exploratory time on the EVLA, in operation since March
2010, to follow up with a confirming observation on one of the
candidate masers in July 2010. By this time 25 antennas were equipped
with the new, final 4-8 GHz C band receivers of which typically 23
produced usable data. Apart from, e.g., the new receivers and receiver
range of the EVLA, the EVLA also utilizes the new WIDAR
correlator. Our EVLA observations used one of the standard modes
available during the WIDAR commissioning period, using 2 MHz (90
km\,s$^{-1}$) bandwidth with 256 dual polarization
channels (7.8 kHz, 0.35 km\,s$^{-1}$). In total 8 hours were spent to
observe the candidate maser over four runs around the end of July
2010, yielding about 1.5 mJy\,beam$^{-1}$ rms noise per channel in the
robust weighted image, but with much narrower channel separation
compared to the earlier VLA observations. Doppler \emph{setting} was
used, where the observing frequency was calculated to center the
observation at $-$240 km\,s$^{-1}$ at the start of the observation and
then kept fixed. This contributed a negligible spectral broadening of
less than 0.1 km\,s$^{-1}$, or about a third of a channel.  The
pointing and frequency were also adjusted compared to our VLA survey to
place the candidate maser closer to the center of the beam and the
center of the observing band.

After confirmation of the methanol maser, we also obtained exploratory
time to search for a potential 22.23508 GHz water maser. The
observation was taken at the end of August 2010, lasted for 2 hours
and used an 8 MHz (100 km\,s$^{-1}$) bandwidth with 256 dual
polarization channels (31.25 kHz, 0.42 km\,s$^{-1}$). The channel
rms noise achieved in the naturally weighted image was 4.8
mJy\,beam$^{-1}$ using data from 25 antennas.

All VLA and EVLA observations were taken when the interferometer was
in its most compact ``{\bf D}'' configuration. The angular resolution
at 6.7 GHz of about 10\arcsec\ corresponds to about 38 pc at the
distance of M31 ($\sim$800 kpc, \citealt[e.g.,][]{stga98}), leaving any candidate maser spatially
unresolved. The 2.8\arcsec\ angular resolution at 22 GHz corresponds
to about 10 pc at the distance of M31. All observations were
exclusively reduced using the standard NRAO Astronomical Imaging
Processing System (AIPS) procedures where a new
Obit\footnote{Available from
http://www.cv.nrao.edu/$\sim$bcotton/Obit.html} task {\tt BDFIn}
filled the native Science Data Model data into AIPS.

\section{Results}

Figure \ref{vladet} shows the primary beam corrected
spectra of the VLA candidate maser and the EVLA maser
confirmation along with a MSX infrared image outlining the location of
the maser in M31. We could identify the methanol
maser in all four EVLA observing chunks (``Scheduling Blocks''), but
the signal-to-noise in the individual scheduling blocks was
insufficient to derive a statement on short time scale
variability. The spectra compare in shape, line width ($\sim$5
km\,s$^{-1}$), peak flux ($\sim$8 mJy\,beam$^{-1}$) and central
line-of-sight velocity ($-$242 km\,s$^{-1}$). Their measured positions
are within 2\arcsec\ (less than one pixel) of each other (at RA
00$^h$44$^m$19.2$^s$, Dec 41$\degr$19\arcmin30\arcsec\ in J2000;
$X$=$+$13\farcm66, $Y$=$+$11\farcm96 following \citealt{baar64}). The
location of the maser is at a projected distance of 18.13\arcmin, 
which corresponds to a deprojected distance of
about 12.7 kpc from the nucleus. No
``VLBI-bright'' water maser complex was found at this location with a
(2$\sigma$) detection limit of 10 mJy\,beam$^{-1}$ per 0.42
km\,s$^{-1}$ channel.

Nearby peaks in CO \citep[near area \#13,][]{dame93} and infrared
radiation (IRAS\,00416$+$4104, 3.17 Jy at 60 $\mu$m) are present.  The narrow velocity range
of the methanol emission is within the wide velocity range of the CO
emission in that region, but a clear relation cannot be derived due to
the low resolution of the CO and infrared data.

\section{Discussion}

We have found and confirmed the first 6.7 GHz methanol maser in M31,
making it the second (and most distant) galaxy outside the Milky Way 
known to host a
methanol maser.  It is also the first maser of any type detected in
M31.

Interstellar methanol is found in massive star forming regions in the
Milky Way, mostly in the spiral arms peaking at around 5 kpc
galactocentric distance, but not being necessarily associated with compact
HII regions \citep{pand10}. Galactic 6.7 GHz methanol masers can be as
bright as a few thousand Jy \citep[e.g.,][]{pest05}. The maser peak
flux density of about 8 mJy\,beam$^{-1}$ for the maser found in M31
compares to masers on the high-end tail of the 6.7 GHz methanol maser
distribution in the Milky Way \citep[see derivation in][]{gold08}. The
brightest Galactic methanol maser with a very accurate distance
measurement is W3(OH): $\sim$3700 Jy at 2.0 kpc \citep{pest05,hach06}.
At this distance, the M31 detection would measure $\sim$1300 Jy, while
the brightest maser in the LMC would be $\sim$3000 Jy
\citep[e.g.,][]{beas96}. Thus, it is plausible that M31 may host
brighter methanol masers than the one reported here.

For the purpose of astrometry and proper motion measurements, higher
frequency masers may be better suited than the here presented 6.7 GHz
methanol maser. For example, a similar but higher frequency Class\,II
methanol maser may be present. A likely candidate would be the 12.2
GHz maser, which can be very bright ($\sim$1100 Jy) but is usually
(though not always) fainter than the brightest 6.7 GHz maser in a
given source \citep{casw95}. 

On the other hand, of the four known 6.7 GHz methanol masers in the
LMC, one was found to have an accompanying water maser. While the LMC
appears to be under-abundant in masers due to its low metallicity
\citep[e.g.,][]{beas96,gree08}, there might be a fair chance of
detecting a water maser near the methanol maser in a metal-rich galaxy
such as M31. The maser detection in M31 is not in the immediate
vicinity ($<$50 pc) of a known star forming region, an H$\alpha$
emission line region, or a compact HII region; regions typically
targeted when searching for water maser emission in the Magellanic
Clouds, M31 and M33.

Assuming that the 6.7 GHz methanol masers are the highest frequency
masers to be found in M31, a conservative estimate of the expected
astrometric accuracy for a low signal-to-noise feature with VLBI
instruments would be about a tenth of the beam size, or about 100
$\mu$as. When we ignore galactic rotation and take the anticipated
lower limit for the transverse motion of M31 of 80 km\,s$^{-1}$
\citep{loeb05,mare08}, which is much higher than an undetermined
random motion of the maser in the galaxy, the expected angular
displacement of M31 would be in the order of 20 $\mu$as\,yr$^{-1}$ (at
800 kpc). This is measurable with about ten years of observations.

A survey for potential similarly bright or brighter methanol masers, as
described in our full survey paper, will allow us to determine
more than one proper motion and derivation of the high-end tail of the
methanol maser distribution. As M31 has a well determined inclination,
the proper motions of a few masers will allow the derivation of an
accurate geometric distance to M31. In that respect it would be very
interesting if these maser proper motions could be linked to a proper
motion measurement of M31*, the $\sim$50 $\mu$Jy\,beam$^{-1}$ (at 6
GHz) nuclear supermassive black hole in M31. We anticipate follow-up
observations, e.g., with the European VLBI Network, for this
purpose. 

\section*{Acknowledgments} 

CEM acknowledges support from the ``Research Experiences for
Undergraduates'' program, which is funded by the National Science
Foundation. The Very Large Array and the Expanded Very Large Array are
operated by the National Radio Astronomy Observatory, which is a
facility of the National Science Foundation operated under cooperative
agreement by Associated Universities, Inc.

{\it Facilities:} \facility{VLA ()}, \facility{EVLA ()}

\end{document}